\documentclass[twocolumn,showpacs,showkeys, amsmath,amssymb,footinbib]{revtex4}
\usepackage{graphicx}
\usepackage{dcolumn}
\usepackage{bm}
\usepackage{epsf}
\newcommand{\be}{\begin{equation}}
\newcommand{\ee}{\end{equation}}
\newcommand{\bea}{\begin{eqnarray}}
\newcommand{\eea}{\end{eqnarray}}
\newcommand{\ba}{\begin{array}{ccc}}
\newcommand{\ea}{\end{array}}
\newcommand{\nn}{\nonumber}

\newcommand{\eq}[1]{eq.~(\ref{#1})}
\newcommand{\eqs}[2]{eqs.(\ref{#1}, \ref{#2})}

\newcommand{\Eq}[1]{Eq.~(\ref{#1})}

\newcommand{\fig}[1]{Fig.\ref{#1}}
\newcommand{\ur}[1]{(\ref{#1})}
\newcommand{\urs}[2]{(\ref{#1},\ref{#2})}
\newcommand{\half}{{\textstyle{\frac{1}{2}}}}
\newcommand{\slk}{\not{\!}}
\newcommand{\slg}{\not{\!\!}}
\def\log{\textnormal{log}}
\def\det{\textnormal{det}}
\def\exp{\textnormal{exp}}
\def\ok{\omega_k}
\def\ge{\gamma_{{\rm E}}}
\def\Tr{ {\rm Tr} }
\def\Sp{{\rm Sp}}

\def\ch{{\rm ch}}
\def\cos{{\rm cos}}

\def\appb#1#2#3{Acta Phys.Polon. B {\bf #1}, #2 (#3)}

\def\ibid#1#2#3{{\it ibid.} {\bf #1}, #2 (#3)}

\def\npb#1#2#3{Nucl. Phys. B {\bf #1}, #2 (#3)}
\def\npps#1#2#3{Nucl. Phys. Proc. Suppl. {\bf #1}, #2 (#3)}

\def\plb#1#2#3{Phys. Lett. B {\bf #1}, #2 (#3)}
\def\pr#1#2#3{Phys.Rev.{\bf #1}, #2 (#3)}

\def\prd#1#2#3{Phys. Rev. D {\bf #1}, #2 (#3)}
\def\prl#1#2#3{Phys. Rev. Lett. {\bf #1}, #2 (#3)}

\def\zpc#1#2#3{Z. Phys. C {\bf #1}, #2 (#3)}
\def\yadf#1#2#3{Yad. Fiz. {\bf #1}, #2 (#3)}
\def\sjnp#1#2#3{Sov. J. Nucl. Phys.  {\bf #1}, #2 (#3)}
\def\rmp#1#2#3{Rev. Mod. Phys.{\bf #1}, #2 (#3)}
\def\jpg#1#2#3{J. Phys. G {\bf #1}, #2 (#3)}
\def\jhep#1#2#3{JHEP {\bf #1}, #2 (#3)}
\def\jetp#1#2#3{JETP Lett. {\bf #1}, #2 (#3)}
\def\pzetf#1#2#3{Pisma Zh.Eksp.Teor.Fiz. {\bf #1}, #2 (#3)}

\begin{document}


\title{Covariant derivative expansion of fermionic effective action at high
temperatures}
\author{Dmitri Diakonov$^{a}$}
\email{diakonov@nordita.dk}
\author{Michaela Oswald$^{b}$}
\email{oswald@alf.nbi.dk}
\affiliation{
$^a$NORDITA,  Blegdamsvej 17, DK-2100 Copenhagen,
Denmark\\
$^b$NBI, Blegdamsvej 17, 2100 Copenhagen, Denmark}
\date{April 30, 2004}

\begin{abstract}
We derive the  contribution of massless fermions to the 1-loop
effective action for  static $A_4$ and $A_i$ fields at high temperatures,
for the $SU(2)$ gauge group assuming that gluon fields are slowly
varying but allowing for an arbitrary amplitude of $A_4$.
\end{abstract}

\pacs{11.15.-q,11.10.Wx,11.15.Tk}
\keywords{gauge theories, finite temperature field theory,
derivative expansion}

\maketitle

\section{Introduction}
Quantum chromodynamics (QCD) at non-zero temperature is an intensely studied
field. At very high temperatures the coupling constant is small and
perturbation theory can be developed. However, due to the chromomagnetic
sector of the theory, perturbation theory explodes already at
a few-loop level \cite{P,L,GPY} and is hence only applicable at
academically high temperatures \cite{breakdown}.
The region of intermediate temperatures is of much bigger interest.
Both the restoration of chiral symmetry and the deconfinement are believed to take
place in this region. In the presence of fermions it is still unclear if there is a
confinement-deconfinement phase transition or just a smooth crossover
between the two phases. In any case QCD is in the deconfined plasma phase
at very high temperatures.

At finite temperature gluons obey periodic and fermions obey anti-periodic boundary
conditions in imaginary time. This property leads to the quantized
Matsubara frequencies. They are even multiples of
$\pi T$ for gluons, i.e. $\ok=2k \pi T$,  and odd multiples of
$\pi T$ for quarks, i.e. $\ok=(2k+1) \pi T$. So while
gluons have a zero mode this is not the case for quarks. This fact has direct and
important influence on the IR behavior of the two contributions to the
effective action.

At the tree level very heavy modes decouple from a theory at high
temperatures. This is called dimensional reduction \cite{dimred}
since the heavy modes are simultaneously the time-dependent
ones. Neglecting all modes except the zero Matsubara frequencies
leaves a 3D static theory
\bea
-\frac{1}{4g^2}F_{\mu\nu}^2+\psi_{f}^{\dagger}i\!\slk{\nabla\psi_{f} \to
}-\frac{1}{4g^2 T}\left[F_{ij}^2+2(D_i^{ab}A_4^b)^2\right],
\label{treelevel}\eea
which only contains the static gluonic  modes with
the coupling constant $g_{\it 3}^2 = g^2 T$. Since the energy can never
vanish for fermions they decouple completely.

The long-range forces mediated by the static gluons
lead to the IR divergencies, because in strict perturbation theory
they are massless. Fermions do not cause any IR problems, since they do not
have zero modes even if they are massless, which is the
case we consider here. Nevertheless, fermions change the effective
action in a drastic way as compared to the pure glue case, since their presence
changes the symmetry of the action with respect to the
center-of-group gauge transformations.

The tree-level action \ur{treelevel} has certainly insufficient
accuracy to study field fluctuations at high but not infinitely high temperatures.
Once one includes quantum corrections both the fermions and all the non-zero
Matsubara modes of the gluons show up in the loops.

Effective theories resulting from quantum corrections for different energy scales,
$T$, $gT$ and $g^2 T$ have been constructed in \cite{coupling,scalecomp}. The
parameters in the effective theories are obtained by matching the correlation
functions between the effective and the actual theory as
functions of the parameters of the original theory. The question of
color-conductivity and transport properties of the plasma has been addressed
in \cite{colorcond}. A heat kernel approach for Yang--Mills theories has been used in
\cite{heatkernel}, a constraint effective potential for the Polyakov loop has
been studied in \cite{CKA} and spatial variations of the Polyakov loop have been
investigated in \cite{fluctpol}.

In the pure Yang-Mills theory the center symmetry plays a crucial part in the
description of the the confinement-deconfinement phase transition
\cite{P,S,'tH,SY}. The latter is usually characterized by an order
parameter which is the average of the trace of the so-called Polyakov line:
\bea\nonumber
P(x)={\rm P}\,\exp\left(i\int_0^{1/T}\!dx_4\,A_4\right).
\eea
The order parameter $<\Tr P>$ is zero in the
confined phase below the critical temperature and assumes a non-zero value in
the deconfined phase above the critical temperature. The Polyakov line is not
invariant under gauge transformations belonging to the gauge group center.
One hence concludes that if $<\Tr P>=0$ then the $Z(N_c)$ symmetry is
manifest. This situation describes confinement. If for any reason $<\Tr
P>\neq 0$ then the symmetry must have been broken. This corresponds to the deconfined phase.

The 1-loop \cite{GPY, WeissYM, WeissF} and 2-loop \cite{2loop} potential
energies as functions of $A_4$ are known. They are periodic functions
of the eigenvalues of $A_4$ in the adjoint
representation with period $2\pi T$. This reflects the symmetry of the
$Z(N_c)$ vacua. The curvature of the potential at its minima gives the
leading order Debye mass for `electric' gluons. The zero energy
minima of the potential correspond to quantized values of $A_4$ or center
group values for the Polyakov line, where $\Tr P\neq 0$. At high
temperatures the system oscillates around one of these minima.
At lower temperatures, however, the fluctuations around the minimum
increase and eventually the system undergoes a phase transition to $<\Tr P>=0$.
At the same time, one expects that near the phase transition point
the fluctuations are long-range.
To study those fluctuations, one needs an effective low-momenta theory
which, however, does not assume that the $A_4$ component is small.

Let us formulate the problem more mathematically. Nonzero temperatures
explicitly break the $4D$ Euclidean symmetry of the theory down to the
$3D$ Euclidean symmetry, so that the spatial $A_i$ and time $A_4$ components
of the Yang--Mills field play different roles and should be treated differently.
One can always choose a gauge where $A_4$ is time-independent.
Taking $A_4(x)$ to be static is not a restriction of any kind on the fields
but merely a convenient gauge choice, and we shall imply this gauge throughout the paper.
[It is also a possible gauge choice at $T=0$ but in that limiting case it is unnatural
as one usually wishes to preserve the $4D$ symmetry.] As to the spatial
components $A_i(x,t)$, they are, generally speaking, time-dependent,
although periodic in the time direction. Putting the components $A_i$
to zero is a gauge non-invariant restriction on the fields since any time-independent
gauge transformation will generate a nonzero $A_i$. Therefore, the spatial derivatives
of the Polyakov line in the gauge-invariant effective action can only
appear as covariant derivatives including a nonzero $A_i$ field.

In \cite{DO} we calculated the 1-loop kinetic energy for the
eigenvalues of the Polyakov line, integrating over gluon and ghost
fluctuations. See also \cite{MO} for a summary. In this work we are
interested in obtaining the effect of quarks on that kinetic energy as well.
We use a background field method for the gluons and evaluate the 1-loop
action through a functional determinant formalism. In particular we assume
the background fields to vary slowly but the  $A_4$ component is allowed
to have an arbitrary amplitude. We integrate out fast varying quantum
fluctuations about them by making an expansion in spatial covariant
derivatives. This method was originally developed in \cite{DPY}
for zero temperature QCD.

This corresponds to summing up all powers of $A_4$ but where their
momenta are restricted to $p<T$ reflecting the long-range behavior of
the plasma phase. As we said, we choose a static gauge for $A_4(x)$.
This gauge choice does not prevent $A_i(x)$ from being time dependent.
Since $A_i(x_4,x)$ is periodic in time, its time derivative is given by the
Matsubara frequencies $\omega_k=2\pi k T$ being $O(T)$ for any $k\neq 0$.
Since we are interested in low momenta fluctuations, $p<T$, it is consistent
to restrict oneself to the zero Matsubara frequency of the background field,
i.e. to the static $A_i(x)$.

We expect that our results are suitable to study the
correlation functions of the Polyakov line not too far from the transition
point where it experiences fluctuations that are
large in amplitude but presumably mainly long ranged.
The results may be of some help for studying quantum
weights of semiclassical objects, such
as dyons (\cite{dyons},\cite{BPS}) or calorons \cite{calorons}.

The effective action contains a contribution from the gluons and from the
fermions. The former part, namely the pure Yang--Mills effective action, was
obtained by the authors in \cite{DO}. Although, as discussed above, there
is no center symmetry for the fermions, it is still instructive to see
their effect on the effective action for the Polyakov line. In particular
we work with the gauge group SU(2), We consider a general
electric field but restrict ourselves to a magnetic field parallel to
$A_4$, $B_i^{\parallel}$. The expected result for the effective
action, which is the sum of the tree-level and 1-loop actions, is hence:
\bea\nonumber
\!\!\!\!\left[S_{{\rm eff}}^{{\rm F}}\right]^{(2)\!\!}&\!\!=\!\!&\!\!
\! \int \frac{d^3 x}{T}  \left[- T^3 V^{{\rm F}}(A_4^2)\!
\qquad\right. \\\label{expect}
&&+ \left. \! E_i^2 F_1^{{\rm (F)}}(A_4^2)\!+\! \frac{(E_i A_4)^2}{A_4^2}
F_2^{{\rm (F)}}(A_4^2)\!\qquad\right. \\
&&+ \left. \! (B_i^{\parallel})^2\,H_1^{{\rm (F)}}(A_4^2)+ \ldots \right]\,,
\nn
\eea
with the electric and magnetic fields
\bea
E_i^a&=&D_i^{ab}A_4^b-\dot A_i^a=\partial_iA_4^a+\epsilon^{acb}A_i^cA_4^b-\dot A_i^a,\\
B^a_i&=&\frac{1}{2}\epsilon_{ijk}\left(\partial_jA_k^a-\partial_kA_j^a
+\epsilon^{abc}A_j^bA_k^c\right).
\eea
The first term is the potential energy and the remaining terms are the
kinetic energy contributions in the color-electric and color-magnetic
sector. The objective of this paper is to find these functions.

\section{The QCD action at finite temperature}
The basics about Yang--Mills theory at finite temperature were discussed in \cite{DO}.
The (Yang--Mills) action of gluons at finite temperature is given by
\bea
S^{{\rm YM}} = \int_0^{\beta} \!\!\!dx_4\int \!\! d^3 x \left[-\frac{1}{4
    g^2(M)}F_{\mu\nu}^a F_{\mu\nu}^a\right],
\quad\beta=\frac{1}{T},
\eea
where the gluon fields obey periodic boundary conditions in the temporal direction,
i.e.
\be\label{per}
A_{\mu}(0, x) = A_{\mu}(\beta, x).
\ee
Because of the compactified time direction there is a group of special gauge
transformations which transform the gluon fields in the usual way as
\be
A_{\mu} \to U\,A_{\mu}\,U^{-1} + i\, U \partial_{\mu}U^{-1}
\ee
and which preserve the periodicity condition \ur{per}, but which are not periodic themselves:
\be\label{antiperU}
U(0, x) = z_k U(\beta, x).
\ee
Here $z_k$ is an element of the center group $Z(N_c)$:
\be
z_k = e^{2\pi i k/N_c} \qquad k\in\{0, N_c-1\}.
\ee
The Yang--Mills action is invariant under this gauge transformation, but the
Polyakov line is not. It transforms as
\bea
P(x)={\rm P}\,\exp\left(i\int_0^{1/T}\!dt\,A_4\right) \to z_k^{-1}P(x).
\eea
From this property one sees immediately that the $Z(N_c)$ symmetry implies
$<\Tr P>=0$ while it must be (spontaneously) broken if $<\Tr P>\neq 0$.

Including $N_f$ quarks with mass $m_f$ the full QCD action at finite temperature becomes
\bea\label{SQCD}
\lefteqn{S= \int_0^{\beta} \!\!\!\!dx_4\int d^3 x} \\ \nonumber && \times
  \left[-\frac{1}{4 g^2(M)}F_{\mu\nu}^a F_{\mu\nu}^a + \sum_{f=1}^{N_f}
    \psi_{f}^{\dagger}\left(i\!\slk{\nabla} + i m_f\right)\psi_{f}\right],
\eea
where the Dirac operator is given by
\be\label{dirac}
i\!\slk{\nabla} = i\!\slk{\partial} +  T^a A_\mu^a\gamma^\mu.
\ee
Here the $T^a$ are the generators of $SU(N)$ in their fundamental
representation, they are half the Pauli matrices for SU(2).
The fermions in \eq{SQCD} obey anti-periodic boundary conditions
\be
\psi_{f}(0, x) = - \psi_{f}(\beta, x).
\ee
This property is, however, not preserved by the $Z(N_c)$ gauge transformation
\ur{antiperU}. Specifically the fermions transform as
\bea
\psi_{f}^{U} (\beta, x) &=& U(\beta, x)\,\psi_{f}(\beta, x),\\ \nonumber
\psi_{f}^{U} (0, x) &=& U(0, x)\,\psi_{f}(0, x) = z_k\, U(\beta, x)\,\psi_{f}(0, x)\\ \nonumber
&=& \!\!- z_k \,U(\beta, x)\,\psi_{f}(\beta, x) = \!\!- z_k\,\psi_{f}^{U} (\beta, x).
\eea
Hence in the presence of fermions the $Z(N_c)$ symmetry gets explicitly broken and
the Polyakov line ceases to serve as an exact order parameter for the theory,
since $<\Tr P>\neq 0$ for all temperatures. Nevertheless, even in the presence 
of massless fermions the Polyakov line might provide useful information near 
the critical temperature \cite{Karsch}.

\section{One loop quantum action}
The partition function of QCD in its Euclidean
invariant form is given by
\bea
\lefteqn{Z(A,\psi_{f}, \bar{\psi_{f}}) = \sum_{f=1}^{N_f}\int DA\, D\psi_{f}\,
  D\psi_{f}^{\dagger}\, \exp\int d^4 x}\\ \nonumber &&\times
   \left[-\frac{1}{4 g^2(M)}F_{\mu\nu}^a F_{\mu\nu}^a + \sum_{f=1}^{N_f}
    \psi_{f}^{\dagger}\left(i\!\slk{\nabla} + i m_f\right)\psi_{f}\right].
\eea
We use the background field method for the gluon fields, where we decompose
them into background fields and quantum fluctuations around them which we
assume to be small:
\be
A_{\mu} = \bar{A}_{\mu}+ a_{\mu}.
\ee
In this work we are interested in a 1-loop effective theory for the
background $\bar{A}$ fields. This
corresponds to an expansion of the action around the background gluon fields to
quadratic order in the quantum fluctuations $a_{\mu}$.
The one loop expansion of the gluon Lagrangian is:
\bea\label{Fexp}
-\frac{1}{4 g^2(M)} F_{\mu\nu}^2(A)&=& -\frac{1}{4 g^2(M)} F_{\mu\nu}^2(\bar A)
\\ \nonumber
&& - \frac{1}{g^2(M)}D_{\mu}(\bar A)F_{\mu\nu}(\bar A)\,a_{\nu}
\\ \nonumber
&&
- \frac{1}{2 g^2(M)}a_\mu^a\;W_{\mu\nu}^{ab}\;a_\nu^b+\ldots
\eea
where
\bea\label{Wdeg}
W_{\mu\nu}^{ab}\! =\! -\left[D^2\!(\bar A)\right]^{ab}\delta_{\mu\nu}\!
+\!\left[D_\mu D_\nu\right]^{ab}\!\!-\! 2f^{acb}F_{\mu\nu}^c(\bar A) ,
\eea
and
\be
D_\mu^{ab}(\bar{A}) = \partial_{\mu}\delta^{ab} + f^{acb}\bar{A}_{\mu}^{c}
\ee
is the covariant derivative in the background field in the adjoint
representation. The second term in \eq{Fexp}, which is linear in $a_{\nu}$,
is zero if the background field obeys the equation of motion.
In the fermionic Lagrangian the quarks couple to the gluon fields in the
usual minimal, i.e. linear way. Hence the expansion is just
\bea\label{Dexp}
\psi_{f}^{\dagger}i\!\slk{\nabla}\psi_{f}
=\psi_{f}^{\dagger} i\!\slk{\nabla}(\bar{A})\psi_{f} + \psi_{f}^{\dagger}a_{\mu}\gamma_{\mu}\psi_{f} +\ldots
\eea
where $\slk{\nabla}(\bar{A})$ is the covariant derivative of the
background field in the fundamental representation. The second term in
\eq{Dexp} contributes at the 2-loop level which we do not consider here.

The quadratic form $W_{\mu\nu}^{ab}$ in \eq{Wdeg} is degenerate: it has an infinite number of zero modes
which are the infinitesimal gauge transformations
$a_\mu^a=D_\mu^{ab}\Lambda^b$. In order to remove this degeneracy one has to
fix the gauge for these fluctuations. We choose the background Lorenz gauge
$D_{\mu}(\bar{A})a_{\mu}=0$ \footnote{Jackson and Okun \cite{JO} recommend
to name the $\partial_\mu A_\mu=0$ gauge after the Dane Ludvig Lorenz
and not after the Dutchman Hendrik Lorentz who certainly used this gauge too
but several decades later.}.
The second term in \eq{Wdeg} cancels out but the Faddeev--Popov ghost
determinant arises which again can be expressed as a Grassmann integral over
ghost fields.

The 1-loop partition function thus becomes
\bea
\lefteqn{Z(\bar{A}) = e^{\bar{S}}\,
\int{}Da\,D\chi\,D\chi^{+}\,D\bar{\psi_{f}}\,D\psi_{f}\,\exp \left\{\int{}d^4 x
\right.}\\\nonumber
&&\!\!\!\!\!\!\left.\times\left[
-\frac{1}{2g^2(M)}\left(a_{\mu}^b\, W_{\mu\nu}^{bc}\, a_{\nu}^c
-\chi^{+a}\,D^2_{\mu}\,\chi^{a}\right)
+ \sum_{f=1}^{N_f}\psi_{f}^{\dagger}\,i\,{\cal \slg{D}}\,\psi_{f}\right]\right\}\;,
\eea
where $\chi, \chi^+$ are ghost fields,
\be
i\,{\cal \slg{D}}_f=i\!\slk{\nabla}+i m_f = i\!\slk{\partial} + T^a
\bar{A}_\mu^a\gamma^\mu + im_f\,
\ee
is the massive Dirac operator in the fundamental representation, and
\be
 \bar{S} =  -\frac{1}{4 g^2(M)}\int\,d^4 x{}F_{\mu\nu}^{a}(\bar{A}) F_{\mu\nu}^{a}(\bar{A})
\ee
is the action of the background gluon fields.

Integrating out the quarks, ghosts and the quantum fluctuations of the gluons
leaves us with the desired effective theory for the background ${\bar A}$
fields:
\be
Z(\bar{A}) = e^{\bar{S}}\; \left(\det{W}\right)^{-1/2} \; \det\left(
  -D^2\right)\; \prod_{f=1}^{N_f}\det\left(i\,{\cal \slg{D}}_f\right)\;,
\ee
so that the 1-loop action is
\bea\label{Seff}
S_{\rm{ 1-loop}} &=& \log\, \left(\det{W}\right)^{-1/2}+ \log\,\det
\left(-D^2\right)\\ \nonumber
&+& \sum_{f=1}^{N_f}\log\,\det\left(i\,{\cal \slg{D}}_f\right)\;.
\eea
Since the $\bar{A}$ are the only gluon fields left we will omit the bar from
now on. So far the background field has been kept arbitrary. One has, however, 
the gauge freedom to choose the $A_4(x)$ fields to be static. The spatial gluon
components are generally time dependent. Since $A_i(x_4,x)$ is periodic in time, 
its time derivative is given by the Matsubara frequencies $\omega_k=2\pi k T$ 
being $O(T)$ for any $k\neq 0$. Since we are interested in low momenta fluctuations, 
$p<T$, we shall restrict ourselves to the zero Matsubara frequency 
of the background field, i.e. to the static $A_i(x)$. 

The operators in the ghost and gluon functional determinants, $D^2$ and $W$,
are matrices in the adjoint representation of the color group, and they are
built from covariant derivatives and the field strength only. We used this
fact in \cite{DO} to make an expansion of the 1-loop pure Yang--Mills action in
powers of $D_i$. Since the (static) electric field is given by 
$E_i^a = D_i^{ab}A_4^b$ and the magnetic field by
$B_k^a=\frac{1}{2}\epsilon_{ijk}F_{ij}^a
=\frac{1}{4}\epsilon_{ijk}\epsilon^{cad}\left[D_i,D_j\right]^{cd}$ 
we obtained an effective action for the background $A_4$ fields in 
terms of electric and magnetic fields.

In this paper we study the contribution of the fermion functional
determinant to that effective action. We use again the technique
of the covariant derivative expansion. In addition, we will work
in the chiral limit throughout, i.e. we set $m_f=0$. The main
difference to the gluon calculation is that in the case of
fermions we are dealing with operators in the fundamental
representation and that we do not expect $Z(N_c)$ symmetric
results.

\section{The  fermionic functional determinant}

Throughout this paper we will be working with Euclidean coordinates. A summary of our
conventions is given in the Appendix. In particular we use the following:
\bea\label{conv}
\{\gamma^\mu, \gamma^\nu\} &=& 2 \delta^{\mu\nu}{\bf{1_{4}}}, \\ \label{sigma}
\left[\gamma^\mu, \gamma^\nu\right] &=& 4\,{\rm i}\sigma^{\mu\nu},
\eea
where the $\gamma_\nu$ denote the Euclidean {\it Dirac} matrices, and
$\sigma^{\mu\nu}$ are the spin matrices.  Since we are
working in the chiral limit the Dirac operator is given by \eq{dirac}
and is by definition hermitian:
\be
i\!\slk{\nabla} = i\!\slk{\partial} +  T^a A_\mu^a\gamma^\mu =
\left(i\!\slk{\nabla}\right)^{\dagger}.
\ee
The covariant derivative defines the field strength tensor in the
fundamental representation as
\be\label{F}
[\nabla_\mu,\nabla_\nu]=-i F_{\mu\nu}\,.
\ee
The functional determinant of the fermions can be written as
\be
\det(i\!\slk{\nabla}) = \sqrt{\det(i\!\slk{\nabla})(i\!\slk{\nabla})}
\ee
which following a method originally introduced by Schwinger \cite{Schw}
can be further expressed as
\bea\label{det1}
\det(i\!\slk{\nabla}) = \exp\left( -\frac{1}{2}\,{\rm Sp}
\int_0^{\infty}\frac{ds}{s} \,e^{-s(i\not{\nabla})(i\not{\nabla}) }\right)\,.
\eea
Here Sp is the functional trace.
For its contribution to the effective action we have to properly
normalize it, i.e. subtract the free zero-gluon part:
\be\label{detn}
\log\,\det(i\!\slk{\nabla})_n = -\half{\rm Sp}\int_0^{\infty}\frac{ds}{s}
\left(e^{s\not{\nabla}^2} - e^{s\not{\partial}^2}\right)\,.
\ee
The square of $\slk{\nabla}$ can be decomposed further,
\be \nonumber
\slk{\nabla}^2 = \gamma_\mu \gamma_\nu \nabla_\mu \nabla_\nu
= \frac{1}{2}\left(\{\gamma_\mu, \gamma_\nu\}\nabla_\mu \nabla_\nu
+ \left[\gamma_\mu, \gamma_\nu\right]\nabla_\mu \nabla_\nu \right)\,.
\ee
Since in the second term on the l.h.s. the commutator is antisymmetric
we can also antisymmetrize
\be\nonumber
\nabla_\mu \nabla_\nu \to\frac{1}{2}\left[\nabla_\mu,\nabla_\nu\right]\,.
\ee
With Eqs. (\ref{conv},\ref{sigma},\ref{F}) one finds
\be
\slk{\nabla}^2 = \nabla^2{\bf{1_{4}}}+ \sigma_{\mu\nu}F_{\mu\nu}\,.
\ee
Equation (\ref{detn}) hence becomes
\bea\label{logdiv}
\lefteqn{\log\,\det(i\!\slk{\nabla})_{n} =
-\frac{1}{2}\,{\rm Sp}\int_0^{\infty}\frac{ds}{s}} \\ \nonumber
&& \times \left\{\exp\left[s(\nabla^2{\bf{1_{4}}}+ \sigma_{\mu\nu}F_{\mu\nu})\right]
- \exp\left[s\,\partial^2{\bf 1_4 }\right]\right\}\,.
\eea

The 1-loop action is UV divergent. This comes from the fact that
the running coupling constant is divergent at the tree level.
Since QCD is a renormalizable theory, the tree level divergence
has to be canceled by a 1-loop divergence. In order to control
the divergent behavior of (\ref{logdiv}) we regularize the determinants
by introducing a Pauli-Villars cutoff M in momentum space.
This means that we use the so-called ``quadrupole formula'':
\bea
&& \det(i\!\slk{\nabla})_{n, r} = \sqrt{\frac{\det(- \slk{\nabla}^2)}
{\det(-\slk{\partial}^2)}\frac{\det(-\slk{\partial}^2 + M^2)}
{\det(-\slk{\nabla}^2 + M^2)}} \\ \nonumber
&&
 = \exp\left\{-\half\int_0^{\infty}\frac{ds}{s}\Sp\,\left[\left(1-e^{sM^2}\right)
\left(e^{s\not{\nabla}^2} - e^{s\not{\partial}^2}\right)\right]\right\}.
\eea
The functional trace $\Sp$ can be taken by inserting
any complete basis. We choose the plane wave basis:
\be
{\rm Sp}\, e^{-s K} = \Tr\int d^4 x \lim_{y\to x}\int\frac{d^4 p}{(2 \pi)^4}\,
e^{-ip\cdot y}e^{-s K}e^{ip\cdot x},
\ee
where $\Tr$ is the remaining matrix trace over color and Lorentz
indices. One can now drag the latter plane-wave exponent though
the differential operator $K$ until it cancels with the former.
This results in the shift of the derivatives inside the
differential operator and in the following representation of
the functional trace \cite{DPY}:
\be\label{shift}
{\rm Sp}\, e^{-s K} =  \Tr \int  d^4 x  \int \frac{d^4 p}{(2 \pi)^4} e^{-s K (\partial_{\alpha}
\rightarrow \partial_{\alpha}+ i p_{\alpha})}\bf{1}\,.
\ee
The $\bf{1}$ at the end is meant to emphasize that the shifted operator acts on unity,
so that for example any term that has a $\partial_{\alpha}$ in the exponent
and is brought all the way to the right, will vanish.
According to ({\ref{shift}) we now have
\bea\label{logdet1}
\lefteqn{\log\,\det(i\!\slk{\nabla})_{n, r}\!= -\frac{1}{2}\!\int \!\! d^3
  x\!\! \sum_{k=-\infty}^{\infty}\!\int\!\! \frac{d^3 p}{(2 \pi)^3}\!\!
  \int_0^{\infty}\!\frac{ds}{s}\!\left(1\!-\! e^{-sM^2}\!\right)}\nonumber
\\ \nonumber
&&\!\!\!\!\!\! \times
\Tr\left\{\exp\left[s(\nabla_4+i\ok)^2{\bf{1_{4}}}+s(\nabla_i+ i p_i)^2{\bf{1_{4}}}
+ s\sigma_{\mu\nu}F_{\mu\nu}\right]\right.\\
&& \!\!\!\!\!\! \left.
- \exp\left[{s(i\ok)^2{\bf{1_{4}}} + s(ip_i)^2}{\bf{1_{4}}}\right]
\right\}\,.
\eea
Let us now define
\be
{\cal B} \equiv \nabla_4+i\ok {\bf 1_2}\,,
\ee
then \eq{logdet1} becomes
\bea\label{logdet}
&&\log\,\det(i\!\slk{\nabla})_{n, r} = -\frac{1}{2}\int\,d^3
  x\sum_{k=-\infty}^{\infty}\int\frac{d^3 p}{(2
    \pi)^3} \\ \nonumber
&&
\times\int_0^{\infty}\frac{ds}{s}\Tr e^{-sp^2}\left(1-
  e^{-sM^2}\right)\left\{- e^{-s\ok^2}\right. \\ \nonumber
&&
\left. + \exp\left[(s{\cal B}^2 + s\nabla_i^2 + 2 i s p_i\nabla_i){\bf{1_{4}}}
    +  s\sigma_{\mu\nu}F_{\mu\nu}\right]
\right\}.
\eea
This result is independent of the gauge group. In the following we will work
with $SU(2)$. In particular we choose the background $A_4$ fields to be
a) static and b) diagonal, i.e.
\be\label{a4diag}
A_4(x) = \phi(x)\frac{\tau_3}{2},
\ee
then
\be\label{B}
 {\cal B} = -i \phi(x)\frac{\tau_3}{2} + i \ok {\bf 1_2}\,.
\ee
Here $\tau_3$ is the third of the three Pauli matrices:
\bea
\tau_1 = \left(
\begin{array}{cc}
0&1\\
1&0
\end{array}\right)\,,
\,\,\,
\tau_2 = \left(
\begin{array}{cc}
0&-i\\
i&0
\end{array}\right)\,,
\,\,\,
\tau_3= \left(
\begin{array}{cc}
1&0\\
0&-1
\end{array}\right) .
\eea

\section{Covariant derivative expansion}
\subsection{Zeroth order - The fermionic potential}
In this order we set the spatial covariant derivative to zero,
i.e. $\nabla_i=0$, but sum over all powers of $\nabla_4$:
\bea
\!\!\left[\log\,\det(i\!\slk{\nabla})_{n, r}\right]^{(0)}\!\!
&=&\!\! -\frac{1}{2}\int\!\!  d^3 x\!\! \sum_{k=-\infty}^{\infty}\!\!\!\!
\int\frac{d^3 p}{(2 \pi)^3}\!\! \int_0^{\infty}\frac{ds}{s} \\ \nonumber
&&\!\!\times \Tr  e^{-s p^2} \left(e^{s{\cal B}^2{\bf{1_{4}}}} - e^{s \ok^2 {\bf{1_{4}}}}\right)\,.
\eea
This can be evaluated explicitly. With \eq{B} it is easy to check that
\be\label{trB}
\Tr e^{s{\cal B}^2{\bf{1_{4}}}}= 4 \,\left[e^{-\frac{1}{4}s(\phi-2\ok)^2}
+e^{-\frac{1}{4}s(\phi+2\ok)^2}\right]\,,
\ee
where the factor $4$ comes from $\Tr{\bf{1_{4}}}$.
Since the fermionic energies are given by $\ok = (2 k+1) \pi T$ we can rewrite the terms in \eq{trB} as
\be\label{fsum}
e^{ - \frac{1}{4}s(2\ok\pm\phi)^2}\equiv e^{- s[2\pi T k - \phi_{\pm}]^2},
\ee
where we defined
\be
\phi_{\pm}=\frac{\phi}{2}\pm\pi T\,.
\ee
The summation over $\ok$ and the integrations over $s$ and $p$ can now be
performed along the lines of the bosonic case (\cite{DO}), using the formula
\be\label{key1}
\!\!\!\!\sum_{k=-\infty}^{\infty}\!\!\int_0^{\infty} \!\!
\frac{ds}{s} e^{-s (2\pi T k - \tilde{\phi})^2 - s p^2} \!=\!
-\log\left(\!\ch\frac{|\vec{p}|}{T}\! -\! \cos\frac{\tilde{\phi}}{T}\!\right)\!.
\ee
The result is the following:
\bea\label{fpot}
&& \left[\log\,\det(i\!\slk{\nabla})_{n, r}\right]^{(0)}
 = \frac{1}{12\pi^2 T} \int d^3 x
\\ \nonumber
&& \times
\left[\phi_{+}^2(2\pi T - \phi_{+})^2 + \phi_{-}^2
(2\pi T - \phi_{-})^2 -  2(\pi T)^4\right]_{{\rm mod}\,\,2\pi T }
\\ \nonumber &&
 = \frac{1}{96\pi^2 T} \int  d^3 x \left\{\left[\phi^2-(2\pi T)^2\right]^2
 - (2\pi T)^4\right\}_{{\tiny{\rm mod}\,\,4\pi T }}\,.
\eea
This potential is symmetric around $\phi=0$ and is periodic
with period $4\pi T$ in contrast to the gluon potential (\cite{GPY},\cite{WeissYM})
which has period $2\pi T$.
The curvature around $\phi=0$ of \eq{fpot} gives the fermionic contribution
to $-m_D^2/T$. Indeed we find $- (N_f\,T)/6$ which is in accordance with the
known \cite{GPY} 1-loop result for the Debye mass
\be\label{debye}
m_D^2 = \frac{1}{3}T^2\left(N_c+\frac{N_f}{2}\right)\,.
\ee
If we add the corresponding result from the gluons (see e.g. \cite{DO})
\be
\frac{1}{12\pi^2 T}\int\,d^3 x\left[\phi^2(2\pi T - \phi)^2\right]|_{{\rm mod}\,\,2\pi T }
\ee
then we get the full result for two colors, namely $(2+N_f/2)\,T/3$.
Introducing the variable $\nu=\phi/(2\pi T)$ we have
$\phi_{\pm}=\pi T(\nu\pm 1)$ and \eq{fpot} becomes
\be
\left[\log\,\det(i\!\slk{\nabla})_{n, r}\right]^{(0)}
= T^3 \frac{\pi^2}{6}\int \left[d^3 x (1-\nu^2)^2 -1\right]_{{\rm mod}\,\,2}\,.
\ee
The potential is then given by the (in the chiral case identical) contributions of
all $N_f$ quark flavors:
\be\label{VF}
V^{{\rm F}} = - N_f\,\frac{(2\pi)^2}{24}\left[(1-\nu^2)^2 -1\right]_{{\rm mod}\,\,2}\,.
\ee
This result is of course well known and can for example be found in the
Appendix D of \cite{GPY} or in \cite{WeissF}.
It can be compared to the pure Yang--Mills potential (\cite{GPY},\cite{WeissYM})
\be\label{VYM}
V^{{\rm YM}}=\frac{(2\pi)^2}{3}\nu^2 (1-\nu)^2|_{{\tiny{\rm mod}\,\,1}}.
\ee
Both potentials are shown in Fig. \ref{Pot}.

\begin{figure}[t]
\centerline{
\epsfxsize=0.4\textwidth
\epsfbox{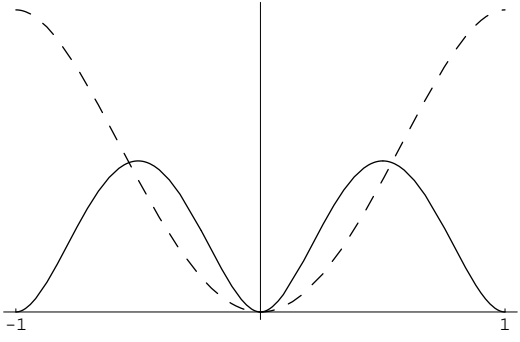}}
\caption{The gluon (solid) and fermion (dashed) potentials for $N_f=1$ and $-1\leq \nu  \leq 1$.} \label{Pot}
\end{figure}

We see clearly that the YM potential has $N_c$ minima, which are the
$Z(N_c)$ symmetric points of $A_4=0$. Since there is no center symmetry for
the fermions, we do not find the same situation. Indeed the fermion potential
has a minimum at $\nu=0$ and a maximum at $\nu=1$. Its period is doubled
relative to the YM potential. This fact comes solely from the
fundamental representation of the fermions.

\subsection{Leading terms in electric sector}

For an effective theory we are interested in the leading terms in the
electric sector.  We would like to stress that we are keeping all
powers of the background $A_4$ field in our approach, but make an
expansion in the spatial covariant derivative. This means that we
allow for an arbitrary amplitude of the $A_4$ fields but we assume
that all the background fields are slowly varying and have momenta $p<T$.
For the leading terms we hence expand to quadratic  order in $\nabla_i$.
Just as for the gluons and ghosts \cite{DO} the technique is to expand
\be\label{expd}
\Tr\,\exp\left\{s\left[\left({\cal B}^2 + \nabla_i^2+2ip_i \nabla_i\right){\bf{1_{4}}}
+ \sigma_{\mu\nu}F_{\mu\nu}\right]\right\}
\ee
in powers of $\nabla_i$ using the following two master formulas for two non-commuting
matrices $A$ and $B$:
\bea\label{master1}
e^{A+B} &=& e^A + \int_0^1\,d\alpha\,e^{\alpha A}B e^{(1-\alpha) A} \\\nonumber
&&+ \int_0^1\,d\alpha\,\int_0^{1-\alpha}\,d\beta\,e^{\alpha
A}B\,e^{\beta A} B e^{(1-\alpha-\beta)}A + \ldots \;
\eea
and
\be\label{master2}
\left[ B, e^A\right] = \int_0^1 d\gamma\,e^{\gamma A}\left[ B,A \right]\,e^{(1-\gamma)A}\;.
\ee
Since in \eq{master1} powers of $B$ are brought down in the expansion we
identify $B$ with the combinations of covariant derivatives in \eq{expd} and $A$
is the rest. The electric field is identified as
\bea\label{elf}
[\nabla_i, {\cal B}] = [\nabla_i, \nabla_4] = -i F_{i4} = -i E_i = - i E_i^a\,T^a.
\eea
To the second order in $\nabla_i$ there are three terms contributing:
\bea
\label{T1}
T_1&=& s \,\Tr \int_0^1 d\alpha\,\,e^{\alpha s {\cal B}^2{\bf{1_{4}}}}
\left(\nabla_i^2 {\bf{1_{4}}}\right) e^{(1-\alpha) s {\cal B}^2{\bf{1_{4}}}} ,\\
\label{T2}
T_2 &=& -\frac{4p^2 s^2}{3} \Tr \int_0^1 d\alpha  \int_0^{1-\alpha}  d\beta\,
e^{\alpha s {\cal B}^2{\bf{1_{4}}}}(\nabla_i{\bf{1_{4}}})
\\ \nonumber && \times\,\,
e^{\beta s {\cal B}^2{\bf{1_{4}}}}\,(\nabla_i{\bf{1_{4}}})\,
e^{(1-\alpha - \beta) s {\cal B}^2{\bf{1_{4}}}} , \\
\label{T3}
T_3 &=&  s^2\, \Tr \int_0^1 d\alpha  \int_0^{1-\alpha}  d\beta\,
e^{\alpha s {\cal B}^2{\bf{1_{4}}}} \,( \sigma_{\alpha\beta}F_{\alpha\beta})
\\ \nonumber && \times\,\,
e^{\beta s {\cal B}^2{\bf{1_{4}}}}\,(\sigma_{\gamma\delta}F_{\gamma\delta})\,
e^{(1- \alpha - \beta) s {\cal B}^2{\bf{1_{4}}}}.
\eea
The terms $T_1$ and $T_2$ are of the same structure as in the gluon
case and can be shown (see Appendix B of \cite{DO}) to yield two gauge
invariant contributions:
\bea\label{I1}\nonumber
I_1 \!\!\! &=&  \!\!\!s^3\int_0^1
d\alpha\,\left\{-\frac{1}{2} + \alpha(1-\alpha)+\frac{2}{9}s
  p^2\left[1-\frac{3}{2}\alpha(1-\alpha)\right]\right\}
\\  && \times
\Tr \,e^{(1-\alpha)
s{\cal B}^2}\left\{{\cal B}, E_i\right\}\, e^{\alpha s{\cal B}^2}
\left\{{\cal B}, E_i\right\}{\bf{1_{4}}}  \;,\\I_2 \!\!\!
&=& \!\!\! -s^2\left( \frac{1}{2}-\frac{2}{9}sp^2 \right)\!\!\Tr e^{s{\cal B}^2} \!\!\!
\left(2E_i^2 + i\left\{{\cal B},\left[\nabla_i, E_i\right]\right\}\right){\bf{1_{4}}}
\label{I2} .
\eea
In contrast to the pure Yang--Mills calculation from \cite{DO} the Lorentz
structure yields a factor 4 from $\Tr{\bf{1_{4}}}$ and since we are dealing
with fermions all matrices are in the fundamental representation. The second
term in $I_2$ contains an anticommutator of ${\cal B}$ and the covariant
divergence of the electric field, which is zero if the background field
obeys the equation of motion. We will discuss it separately in the next
section and leave it out for the time being.

What has to be evaluated is
\bea\label{Sel}
\lefteqn{\left[S_{{\rm 1-loop}}^{{\rm F}}\right]_E^{(2)} =
-N_f\,\left[\log\,\det(i\!\slk{\nabla})_{n, r}\right]_E^{(2)}}\\ \nonumber &=&
-\frac{N_f}{2}\int\,d^3 x\sum_{k=-\infty}^{\infty}\int\frac{d^3 p}{(2 \pi)^3}
\int_0^{\infty}\frac{ds}{s} \, e^{-s p^2} \left(I_1+I_2+T_3\right) \\ \nonumber
& \equiv& -N_F\,(L_1 + L_2 + L_3)\label{S3terms}.
\eea
For the summation over the Matsubara frequencies it turns out to be necessary
to define a region of definition for $\phi$. In particular we shall rescale
this field as $\phi=2\pi T \nu$ and look at the interval $-1\le\nu \le 1$. In
different regions of $\phi$ the results will have different functional
forms. We already saw for the fermion potential that it is symmetric in the
interval $-1\le\nu \le 1$, and outside this region one has to continue analytically.

We will start with the term $T_3$. From
\be
\sigma_{\mu\nu}=-\frac{i}{4}[\gamma_{\mu}, \gamma_{\nu}]
\ee
and
\be\label{tr4g}
\Tr\,\gamma_{\alpha} \gamma_{\beta} \gamma_{\gamma} \gamma_{\delta} = 4 (\delta_{\alpha\beta}\delta_{\gamma\delta}
- \delta_{\alpha\gamma}\delta_{\beta\delta} + \delta_{\alpha\delta}\delta_{\beta\gamma})
\ee
it follows  that
\be\label{tr2s2f}
\Tr\,\sigma_{\alpha\beta}\sigma_{\gamma\delta}F_{\alpha\beta}F_{\gamma\delta} = 2\,F_{\alpha\beta}F_{\alpha\beta}\,.
\ee
Since $F_{\alpha\beta}F_{\alpha\beta}$ contributes to the electric sector as $2 E_k E_k$, \eq{T3} is equal to
\bea\label{T3n}\nonumber
T_3 \!\!&=&\!\! 4 s^2\Tr \int_0^1\!\!\! d \alpha\int_0^{1-\alpha}\!\!\! d\beta\,
e^{\alpha s {\cal B}^2}\! E_k  e^{\beta s {\cal B}^2 }\! E_k e^{(1-\alpha-\beta) s {\cal B}^2 }\!\!,\\
\!\!&=&\!\! 2 s^2\Tr \int_0^1\!\! d\alpha e^{\alpha s {\cal B}^2} E_k e^{(1-\alpha) s {\cal B}^2 } E_k,
\eea
where we used in the last line that the integrand is symmetric in $\alpha$
and $\beta$, invariant under $\alpha\to(1-\alpha)$ as well as the cyclicity
of the trace. After integration over $\alpha$, $p$ and $s$ we find the following
structure for $L_3$:
\be\label{L3}
L_3 = \frac{1}{4\pi^2}\left[\left((E_i^1)^2 + (E_i^2)^2\right)\,\frac{\pi}{8}\,S_1+  (E_i^3)^2\,\frac{\pi}{4}\,S_3\right]\,,
\ee
where
\bea\label{S1k}
S_1 \!\! &=&\!\!   \sum_{k=-\infty}^{\infty}\!\! \frac{|\phi
    + 2\ok| - |\phi - 2\ok|}{\phi\ok} = \frac{2}{\pi T} (\log\,4\mu)\qquad \\ \label{S3k}
S_3\!\! &=&\!\! \sum_{k=-\infty}^{\infty}\left(\frac{1}{|\phi - 2\ok|} + \frac{1}{|\phi + 2\ok|}\right)
\\ \nonumber
&=& - \frac{1}{4 \pi T}\left[4\left(\ge-\log\,\mu\right) + \Phi(\nu) \right]\,.
\eea
Here we used that  $\ok=(2 k +1)\pi T$ and $\phi=2\pi T \nu$.
The function $\Phi(\nu)$ is given by
\bea\label{Phi}
\Phi(\nu) = 2
\left[\psi\left(\frac{1+\nu}{2}\right)+\psi\left(\frac{1-\nu}{2}\right)\right].
\eea
Here $\psi$ is the digamma function
\be
\psi(z) = \frac{\partial}{\partial z}\,\log \Gamma(z).
\ee
The parameter $\mu$ is the UV-cutoff in divergent series:
\be
\sum_{k=1}^{\infty}\frac{1}{k}\to\sum_{k=1}^{\mu}\frac{1}{k}\equiv \log\,\mu \,,
\ee
and is related to the Pauli--Villars mass as
\bea
\mu=\frac{M}{4\pi T}e^{\ge}.
\eea
This subtraction scale for the running coupling constant has been known
previously \cite{coupling} and was also obtained in \cite{DO}.

Next we will turn to the invariant $I_1$. After integration over $\alpha$,
$p$ and $s$ we find the following structure for $L_1$:
\be
L_1 = \frac{1}{4\pi^2}\left[\left((E_i^1)^2 + (E_i^2)^2\right)\,\frac{\pi}{24}\tilde{S}_1
+ (E_i^3)^2\frac{\pi}{12}S_{3}(\nu)\right]\,,
\ee
where $S_3$ is given by \eq{S3k} and
\bea\label{tS1k}
\tilde{S}_1 &=& \sum_{k=-\infty}^{\infty}\frac{(\phi^2 + 4\ok^2)}{\phi^3 \, \ok} \quad
\\ \nonumber && \times
\left\{\frac{\phi^2 - 2\phi\ok + 4\ok^2}{|\phi + 2 \ok|}
- \frac{\phi^2 + 2\phi\ok + 4\ok^2}{|\phi - 2 \ok|}\right\}\\
\nonumber
&=& \frac{1}{\pi T}\left[2\left(\log\,\mu - \log\,4 - 2 \ge\right) - \Phi(\nu)\right]\,.
\eea

Finally we investigate $I_2$. Again after integration over $\alpha$, $p$
and $s$, $L_2$ is of the form:
\be
L_2 = - \frac{1}{4\pi^2}\left[\left((E_i^1)^2 + (E_i^2)^2 + (E_i^3)^2\right)\,\frac{\pi}{6}\,S_3\right]\,,
\ee
with $S_3$ given by \eq{S3k}.

Collecting all terms from $L_{1,2,3}$ we find the following results for the
kinetic energy in the electric sector
\bea\nonumber
\!\!\left[S_{{\rm 1-loop}}^{{\rm F}}\right]_E^{(2)}&\!\!\!=&\!\!\!\int\!\! \frac{d^3 x}{T} \left[
  \left((E_i^1)^2 + (E_i^2)^2\right) f_1^{{\rm (F)}}(\nu) + (E_i^3)^2 f_3^{{\rm (F)}}(\nu)\right] \\\label{elkin}
\!\!&\!\!\!=&\!\!\!  \int \!\!\frac{d^3 x}{T} \left[ E_i^a E_i^a  f_1^{{\rm (F)}}(\nu)+ \frac{(E_i^a
    A_4^a)^2}{A_4^b A_4^b}  f_2^{{\rm (F)}}(\nu)\right],
\eea
where $f_2^{{\rm (F)}}(\nu)\equiv f_3^{{\rm (F)}}(\nu)
- f_1^{{\rm (F)}}(\nu)$. In the second line of \eq{elkin} we used
\be
(E_i^3)^2 = \frac{(E_i^a A_4^a)^2}{A_4^b A_4^b},
\ee
which follows from \eq{a4diag}. The functions are given by
\bea\label{k1}
f_1^{{\rm (F)}}(\nu) &=& -\frac{N_f}{24\pi^2}\log\,4\mu,\\ \label{k3}
f_3^{{\rm (F)}}(\nu) &=& \frac{N_f}{96 \pi^2}
\left[4\left(\ge - \log\,\mu\right) + \Phi(\nu)\right], \\ \label{k2}
f_2^{{\rm (F)}}(\nu) &=& \frac{N_f}{96 \pi^2}
\left[4\left(\ge + \log\,4\right) + \Phi(\nu)\right].
\eea
We would like to stress once more that these functions are the results for
the interval $-1\le\nu \le 1$. They are plotted in \fig{k1k2k3} and one sees
that they are symmetric. To get outside the interval $-1\le\nu \le 1$ one has
to continue analytically, and the functional form of the $f_i^{{\rm (F)}}$ changes.

\begin{figure}[t]
\centerline{
\epsfxsize=0.4\textwidth
\epsfbox{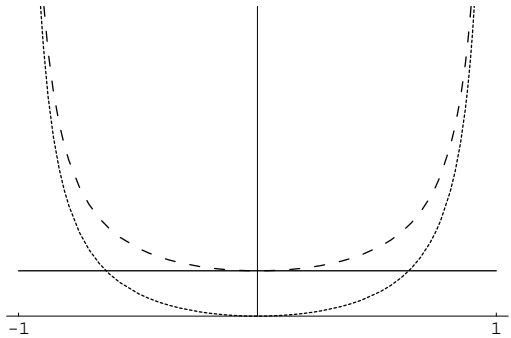}}
\caption{The functions $f_1^{{\rm (F)}}$ (solid), $f_2^{{\rm (F)}}$ (dotted) and $f_3^{{\rm (F)}}$ (dashed)
  without the UV divergent term. \label{k1k2k3}}
\end{figure}
We notice that the function $f_1^{{\rm (F)}}$ is constant, i.e. independent of
$A_4$. However, it contains the UV divergent $\log\,\mu$ that is necessary to
renormalize the running coupling constant from the tree level action:
\be\label{run}
-\frac{F_{\mu\nu}^2}{4 g^2(M)} = -\frac{F_{\mu\nu}^2}{8\pi^2}
\log\,\frac{M}{\Lambda}\left(\frac{11}{12} N_c - \frac{1}{6} N_f\right).
\ee
Here $N_c$ denotes the number of colors and $N_f$ the number of flavors. We
correctly obtained the gluonic contribution to the charge renormalization in
\cite{DO}. For the fermions the tree level divergence in the electric sector
is
\be
\frac{E_k E_k}{2 g^2(M)} = E_k E_k\frac{N_f}{24 \pi^2}\log\,\frac{M}{\Lambda}.
\ee
From our result \eq{k1} we find the correct UV divergent contribution
\be\label{gtree}
-\frac{E_k E_k}{24 \pi^2}\log\,\mu = -E_k E_k \frac{N_f}{24 \pi^2}
\log\,\left(\frac{M}{4\pi T}e^{\ge}\right).
\ee
If we add the tree-level  and the 1-loop action then the
result should be UV finite. This is obtained by choosing the
scale $M$ in \eq{gtree} to be equal to the Pauli--Villars mass, 
which corresponds to the evaluation of the running coupling constant 
at the scale $4\pi T/\exp(\ge)$. In the effective action we then have
to replace the Pauli-Villars cutoff $M$ by $\Lambda$ and find
\bea\label{F1}
F_1^{{\rm (F)}}(\nu) &=& -\frac{N_f}{24\pi^2}\log\,\frac{\Lambda}{4\pi T}e^{\ge},\\ \label{F3}
F_3^{{\rm (F)}}(\nu) &=& \frac{N_f}{96 \pi^2}\left[-4\log\,\frac{\Lambda}{4\pi T}
+ \Phi(\nu)\right], \\ \label{F2}
F_2^{{\rm (F)}}(\nu) &=& \frac{N_f}{96 \pi^2}\left[4\left(\ge + \log\,4\right) + \Phi(\nu)\right]
\eea
with $\Phi(\nu)$ given by \eq{Phi}. 

\subsection{The ``equation of motion'' term}

In the previous section we left away the contribution of the second term in
\eq{I2}. Its contribution to the effective action is
\bea
S_{{\rm EM}} &\!\!=&\!\! \frac{N_f}{2}\int \!\!  d^3 x \!\!
\sum_{k=-\infty}^{\infty} \!\!\int\frac{d^3 p}{(2\pi)^3} \!\!
\int_0^{\infty}\frac{ds}{s} e^{-s p^2} \\ \nonumber
&\!\!\!\!\!\!\times&\!\!\! 4s^2\left( \frac{1}{2}-\frac{2}{9}sp^2 \right)
\Tr e^{s{\cal B}^2} \!\left(i\left\{{\cal B},\left[\nabla_i, E_i\right]\right\}\right).
\eea
After integration and summation this becomes
\bea\label{EM}
S_{{\rm EM}} &=& -\frac{N_f}{12 \pi}\int d^3 x \frac{\Tr\left([\nabla_i, E_i]
  A_4\right)}{\pi T} \\\nonumber
&=&  -\frac{N_f}{12 \pi}\int  d^3 x \frac{(D_i E_i)^a A_4^a}{2\pi T}.
\eea
Here we wrote the result once in terms of a covariant derivative in the
fundamental representation and once through a covariant derivative in the
adjoint representation
\bea
D_i^{ab} = \partial_i \delta^{ab} + f^{acb}A_i^c \quad{{\rm and}}\quad(D_i
E_i)^a = D_i^{ae}E_i^e,
\eea
in order to compare to the gluon results
\cite{DO}.
In \cite{DO} we obtained two terms: one comes solely from the non-zero
Matsubara modes, while the other is the contribution of the zero mode
alone. Our result here, \eq{EM}, is equal to $-N_f/2$ times the first term of
the gluon results.

\Eq{EM} is zero if the background field obeys the equation of motion, $D_i E_i = 0$.
Otherwise it depends on the behavior of $A_4$ and $E_i$ at spatial infinity.
One can integrate \eq{EM} by parts and gets
\bea\label{empart}
S_{{\rm EM}} = \frac{N_f}{24 \pi^2 T} \int d^3 x \left\{E_i^a
    E_i^a - \partial_i\left(E_i^a A_4^a\right)\right\}.
\eea
It yields a contribution to the function \eq{F1} plus a full derivative
term. There are certain background fields, BPS dyons \cite{BPS} being an
example, where $A_4\to$ const. and $E_i \sim \frac{1}{r^2}$ at spatial
infinity. Therefore the full derivative term in \eq{empart} is
nonzero. However, in the particular case of the dyon, it satisfies the
equation of motion and the two terms in \eq{empart} cancel
out.

\subsection{Comparison to previous work}

In a related publication by Wirstam \cite{Wirstam} an effective theory for
QCD at high temperatures was derived by calculating gluon by gluon scattering
at low momenta in terms of Feynman diagrams. In order to compare to the
results of \cite{Wirstam} we have to expand our functions \urs{F1}{F2} to
quadratic order in $\nu$. For \eq{F1} this gives naturally zero, and the
remaining contribution is
\bea
-\int\,d^3 x \frac{7\zeta(3)N_f}{384 \pi^4 T^3}A_4^a A_4^a E_i^b E_i^b,
\eea
which agrees with the result found in \cite{Wirstam} if the gauge group is
chosen to be SU(2).

\subsection{Leading terms in magnetic  sector}
For an effective action in terms of magnetic fields we have to
expand \eq{logdet} to quartic order in $\nabla_i$.
The basic idea of the calculation is, again, to use master equations
(\ref{master1}-\ref{elf}) to drag covariant derivatives $\nabla_i$
to the right. One has for the commutators
\begin{equation}
\nabla_i\,e^{s{\cal B}^2} = e^{s{\cal B}^2} \nabla_i - is\int_0^1\!
d\delta e^{\delta s{\cal B}^2}\{{\cal B},E_i\}e^{(1-\delta) s{\cal B}^2},
\nonumber\end{equation}
\begin{equation}
\nabla_i \nabla_j\,e^{s{\cal B}^2} = e^{s{\cal B}^2} \nabla_i
\nabla_j - is\int_0^1\!\! d\delta e^{\delta s{\cal B}^2}[\nabla_i
\nabla_j, {\cal B}^2]e^{(1-\delta) s{\cal B}^2},
\nonumber\end{equation}
where
\begin{equation}
[\nabla_i \nabla_j, {\cal B}^2] = -i \nabla_i\{{\cal B},E_j\} -
i\{{\cal B},E_i\}\nabla_j.
\nonumber\end{equation}
In this way one ultimately obtains gauge-invariant combinations
of the electric field in the fourth power, mixed terms containing both electric
and magnetic fields, derivatives of the electric field and, finally, magnetic
field squared. In this paper we restrict ourselves to the latter terms
quadratic in the magnetic field $B_i$ defined as
\be
B_i = \half\epsilon_{ijk}F_{jk} = B_i^a\,T^a ,
\ee
where $F_{jk} = i[\nabla_j, \nabla_k]$ in the fundamental
representation. For that reason we shall disregard the commutators $[\nabla_i,A_4]$ as they
introduce powers of $E_i$. In addition, we restrict ourselves to the magnetic 
field parallel to $A_4$, i.e. $B_i=B_i^3\,T^3$. It means that we set the commutator
$[F_{ij}, A_4]=i\,\left([\nabla_i,E_j]-[\nabla_i,E_j]\right)$ to zero. 
In practical terms this means that we can drag all powers of the covariant derivative 
$\nabla_i$  as well as of the field strengths $F_{ij}$ through the exponentials of $A_4$, 
as if they commute. 
Looking at the argument of the exponent in \eq{logdet} we see
that terms which are quadratic in $B_i$ either do not contain the field
strength tensor $F_{\mu\nu}$ at all or consist only of powers of the
latter. Mixing terms vanish upon integration over momentum.
Hence similar to the gluon case (see Appendix C of \cite{DO}) we have to
evaluate the following:
\bea\label{logdetmag}
&&\left[S_{{\rm 1-loop}}^{{\rm F}}\right]_M^{(2)} =
  -N_f\,\left[\log\,\det(i\!\slk{\nabla})_{n, r}\right]_M^{(2)} \\ \nonumber
&&=
 \!\!-\frac{N_f}{2}\!\! \int d^3 x \!\!\sum_{k=-\infty}^{\infty} \!\!
 \int \!\!\frac{d^3 p}{(2 \pi)^3} \!\!\int_0^{\infty} \!\!\frac{ds}{s}
e^{-sp^2}\Tr\,\left[
\left(V_1 \!+ \! V_2\right) \! {\bf{1_{4}}}\right],\\ \nonumber
&& \equiv - N_f\left[N_1 - N_2\right],
\eea
where $V_1$ only contains powers of the derivatives 
\bea\label{V1}
V_1&=&e^{s {\cal B}^2}\left(\frac{s^2}{2}\nabla^2 \nabla^2
+ \frac{(2is)^4}{4!}p_i p_j p_k p_l \nabla_i \nabla_j \nabla_k \nabla_l \right.\\ \nonumber
&+&\left.\frac{(2is)^2 s}{3!}p_i p_j \left[\nabla^2 \nabla_i \nabla_j
+ \nabla_i \nabla^2 \nabla_j + \nabla_i \nabla_j \nabla^2\right]\right) ,
\eea
and $V_2$ comes from the field strength tensor alone:
\bea
V_2 = &&s^2 \Tr\int_0^1 d\alpha \int_0^{1-\alpha} d\beta\,
e^{\alpha s{\cal B}^2}\sigma_{ij}F_{ij}e^{\beta s{\cal B}^2}
\\ \nonumber &&\times\,\,
\sigma_{km}F_{km} e^{(1-\alpha-\beta)s{\cal B}^2}.
\eea
We start with the evaluation of $V_1$. For the momentum integration in we use
\bea\label{d3pint}
\int\frac{d^3 p}{(2\pi)^3} e^{-s p^2} &=& \frac{1}{(4 \pi s)^{3/2}}\; ,\\
\int\frac{d^3 p}{(2\pi)^3}\,p_i\,p_j\, e^{-s p^2}
&=&\frac{1}{2 s} \frac{1}{(4 \pi s)^{3/2}}\,\delta_{ij}\;
,\\
\int\frac{d^3 p}{(2\pi)^3}\,p_i\,p_j\,p_k\,p_m\, e^{-s p^2}
&=&\frac{1}{(2 s)^2} \frac{1}{(4 \pi s)^{3/2}}
\\ \times && \nonumber
\left[\delta_{ij}\delta_{km} + \delta_{ik}\delta_{jm} + \delta_{im}\delta_{jk}\right]\;,
\eea
and obtain the following contribution to \eq{logdetmag}:
\bea\label{N_1}
\!\!\!\!\!N_1 \! = \!\frac{1}{4 \pi^{3/2}}\!\!\!\sum_{k=-\infty}^{\infty}\!
\int_0^{\infty}\!\!\!\frac{ds}{\sqrt{s}}\Tr e^{s{\cal B}^2}\!\frac{1}{12}[\nabla_i,\nabla_j] [\nabla_i,\nabla_j].
\eea
Since $[\nabla_i, \nabla_j]^2 = -F_{ij}F_{ij} = -2 B_k B_k$, where
$B_k=B_k^3\,T^3$, this is equal to
\be
N_1 = -\frac{1}{24 \pi^{3/2}}\sum_{k=-\infty}^{\infty}
\int_0^{\infty}\frac{ds}{\sqrt{s}} \Tr\left( e^{s^2{\cal B}}B_k B_k\right),
\ee
and after integration over $s$ and the summation over the Matsubara
  frequencies it becomes
\be\label{N1}
N_1 = -\frac{1}{24 \pi^{3/2}}\int d^3 x (B_i^3)^2 \frac{\sqrt{\pi}}{2}\,S_3,
\ee
where $S_3$ is again given by \eq{S3k}.
For  $V_2$ we find, using eqs.(\ref{tr4g},\ref{tr2s2f}):
\bea
\!\!\!\!\!V_2\!=\!2 s^2 \Tr\!\!\int_0^1\!\! \!\! d\alpha \!\!
\int_0^{1-\alpha} \!\!\!\!\!\!d \beta e^{\alpha s{\cal B}^2}F_{ij}
\,e^{\beta s{\cal B}^2}F_{ij} e^{(1-\alpha-\beta)s{\cal B}^2}\!.
\eea
Since the contribution to the magnetic sector of $F_{ij}F_{ij}$ is $2 B_k
B_k$ this is equal to
\bea\label{V2}
V_2 &=& 4 s^2 \Tr\!\!\int_0^1\!\! d\alpha \int_0^{1-\alpha}\!\!
d\beta e^{\alpha s{\cal B}^2}B_k e^{\beta s{\cal B}^2}B_k e^{(1-\alpha-\beta)s{\cal B}^2}\quad
\nonumber \\
&=& 2 s^2\,\Tr\,\left(e^{s{\cal B}^2} B_k B_k\right),
\eea
where we used the fact that the integrand is symmetric in $\alpha$ and
$\beta$, the cyclic property of the trace and eventually dragged the magnetic
field to the right. One hence sees that $V_2$ is of the same structure as
$V_1$. Explicitly we find after the integrations over $p,s$ and the
summation over $\ok$ that
\be\label{N2}
N_2 = \frac{1}{16 \pi}\int d^3 x\, (B_i^3)^2\,S_3,
\ee
where $S_3$ is given by \eq{S3k}.
Adding \eqs{N1}{N2} we find the following result for the kinetic energy in
the magnetic sector:
\bea\label{magkin}
\left[S_{{\rm 1-loop}}^{{\rm F}}\right]_M^{(2)}
\int  \frac{d^3 x}{T} \, (B_i^3)^2 \,h_1^{{\rm (F)}}(\nu)\,,
\eea
where the coefficient is given by:
\bea\label{m1}
\!\!h_1^{{\rm (F)}}(\nu) =  \frac{N_f}{96\pi^2}\left[4\left(\ge - \log\,\mu\right)
  + \Phi(\nu)\right]\!,
\eea
with $\Phi(\nu)$ as in \eq{Phi}. 

The function $h_1^{{\rm (F)}}$ above is the result for
$-1\le\nu\le 1$ and it is symmetric in this interval. 
It also contains the necessary UV
divergent contribution to cancel the tree-level divergence of the running coupling constant, \eq{run}:
\be
-\frac{B_k B_k}{24 \pi^2}\log\,\mu = -\frac{B_k B_k}{24 \pi^2}\log\,\left(\frac{M}{4\pi T}e^{\ge}\right).
\ee
Adding up the tree-level and 1-loop terms is obtained by replacing $\mu$ in
\eq{m1}  by $\Lambda/(4\pi T)\exp(\ge)$, which corresponds to an
evaluation of the running coupling constant at the scale $4\pi T/\exp(\ge)$. The final
result is then:
\bea\label{H1}
H_1^{{\rm (F)}}(\nu) = \frac{N_f}{96 \pi^2 T}\left[-4\,\log\,\frac{\Lambda}{4\pi T} 
+ \Phi(\nu) \right].
\eea

\section{Conclusions}

We have calculated the 1-loop contribution of massless quarks to the effective action
at high temperatures for any value of $A_4$ and hence of the Polyakov
line. While we have a general result in the color-electric sector, we
restrict ourselves to a magnetic field parallel to $A_4$. 
The covariant derivative expansion of this action has the form:
\bea\nn
\!\!\!\!\left[S_{{\rm eff}}^{{\rm F}}\right]^{(2)\!\!}&\!\!=\!\!&\!\!
\! \int \frac{d^3 x}{T}  \left[- T^3 V^{{\rm F}}(\nu)+ E_i^2 F_1^{{\rm (F)}}(\nu)\qquad\right. \\
&& \left. 
+ (B_i^{\parallel})^2\,H_1^{{\rm (F)}}(\nu)+\! \ldots\! \right]\,,
\label{final}
\eea
where $\nu =\frac{\sqrt{A_4^aA_4^a}}{2\pi T}$ and $B_i^{\parallel}$ is the magnetic field
parallel in color space to $A_4$. Because of the Bianchi identity,
$[F_{ij}, A_4]=i\,\left([\nabla_i,E_j]-[\nabla_i,E_j]\right)$, in the case where  
the magnetic field is not parallel to $A_4$ one also has to include terms with electric field 
and its derivatives into the effective action, otherwise it will not be complete.

The potential $V^{{\rm F}}$ has double the period as compared to the gluon induced 
potential, is symmetric in $\nu$ between -1 and 1 and has been known before. 
It has its minimum at $|A_4|=0$ and a maximum at $|A_4|= \pm 2\pi T$. The functions
$F_{1,2}^{{\rm (F)}}$ and $H_1^{{\rm (F)}}$ given by \eqs{F1}{F2}
and \eq{H1} are new. All functions, both in the electric and in
the magnetic sector, are symmetric in $\nu$ between -1 and 1, which reflects
the fact, that fermions are in the fundamental representation of the color
group. Our results can be used for studies of QCD at high but
not infinite temperatures, where the Polyakov line experiences fluctuations
which are large in amplitude but long ranged and where the dimensional reduction
approach is too crude. \\


{\noindent\bf Acknowledgments}\\

We are grateful to Chris Korthals-Altes for a critical reading
of the manuscript and for helpful comments.

\vskip 0.5true cm
\noindent{\bf Note added in proof}\\

\noindent
After the submission, the paper by E. Megias, E. Ruiz Arriola and
L.L. Salcedo, {\tt hep-ph/0312133} appeared on the net, in which the authors
compute the
  coefficients in the parallel electric and magnetic sector. Apart from a
  constant our
  functions $h_1^{{\rm (F)}}$ and $f_3^{{\rm (F)}}$ agree with
  their results.

\begin{appendix}
\section{Euclidean coordinates}
In finite temperature QCD one needs the Euclidean formulation of path integrals 
in order to give the partition function the statistical-mechanics interpretation. 
Superscripts $M$ will denote Minkowski coordinates, while superscripts $E$ will refer to
Euclidean coordinates. Note that throughout the paper we have used Euclidean
coordinates without any explicit superscripts.

For space-time coordinates we have
\bea
x_4^E = i x_0^M \qquad x_i^E = x_i^M.
\eea
For the gluon and fermion fields we have
\bea
A_4^E &=& - i A_0^M \qquad A_i^E = A_i^M \\
\psi_{f}^E &=& \psi_{f}^M \qquad \psi_{f}^{\dagger E} = i \bar{\psi_{f}}^M.
\eea
The Dirac gamma matrices are related as:
\bea
\gamma_4^E = \gamma_0^M \qquad \gamma_i^E = -i \gamma_i^M \qquad \gamma_5^E =
\gamma_5^M.
\eea
The Euclidean gamma matrices are hermitian.

\end{appendix}

\end{document}